\documentclass[conference]{IEEEtran}

\usepackage[cmex10]{amsmath}
\usepackage{amssymb}

\usepackage{algorithm, algpseudocode}

\usepackage{graphicx}
\usepackage[capitalise]{cleveref}

\ifCLASSOPTIONcompsoc	\usepackage[caption=false,font=normalsize,labelfont=sf,textfont=sf]{subfig}
\else \usepackage[caption=false,font=footnotesize]{subfig}
\fi

\ifCLASSOPTIONcompsoc \usepackage[nocompress]{cite}
\else                 \usepackage{cite}
\fi

\begin{document}

\title{Trajectory-Aware Rate Adaptation for Flying Networks}

\author{
    \IEEEauthorblockN{
        Ruben Queiros,
        Jose Ruela,
        Helder Fontes,
        Rui Campos
    }
    \IEEEauthorblockA{
        INESC TEC and Faculdade de Engenharia, Universidade do Porto, Portugal \\
        \{ruben.m.queiros, jose.ruela, helder.m.fontes, rui.l.campos\}@inesctec.pt
    }
}
\maketitle

\begin{abstract}
Despite the trend towards ubiquitous wireless connectivity, there are scenarios where the communications infrastructure is damaged and wireless coverage is insufficient or does not exist, such as in natural disasters and temporary crowded events. Flying networks, composed of Unmanned Aerial Vehicles (UAV), have emerged as a flexible and cost-effective solution to provide on-demand wireless connectivity in these scenarios. UAVs have the capability to operate virtually everywhere, and the growing payload capacity makes them suitable platforms to carry wireless communications hardware. The state of the art in the field of flying networks is mainly focused on the optimal positioning of the flying nodes, while the wireless link parameters are configured with default values. On the other hand, current link adaptation algorithms are mainly targeting fixed or low mobility scenarios.

We propose a novel rate adaptation approach for flying networks, named Trajectory Aware Rate Adaptation (TARA), which leverages the knowledge of flying nodes' movement to predict future channel conditions and perform rate adaptation accordingly. Simulation results of 100 different trajectories show that our solution increases throughput by up to 53\% and achieves an average improvement of 14\%, when compared with conventional rate adaptation algorithms such as Minstrel-HT.
\end{abstract}


\section{Introduction}

Even though the concept of ubiquitous wireless connectivity is becoming a reality, there are scenarios where wireless communications coverage is insufficient or does not exist. Considering natural and man-made disaster scenarios, communications infrastructures may be damaged and become unavailable. In temporary crowded events, the existing infrastructure may not have been designed to cope with the additional traffic demand, resulting in overload. In maritime scenarios, environmental monitoring activities using autonomous vehicles will take place in offshore zones, typically not in range of existing onshore communications infrastructures.

\begin{figure}
  \centering
  \includegraphics[width=\linewidth]{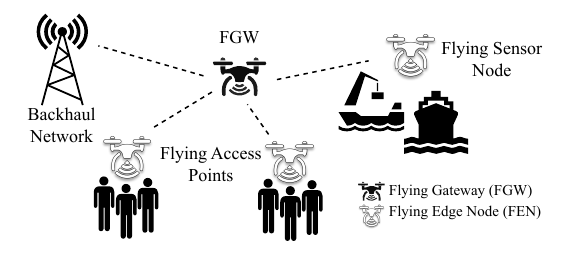}
  \caption{Flying Network multi-tier topology example.}
  \label{fig:flyingNetwork}
\end{figure}

Flying networks, composed of Unmanned Aerial Vehicles (UAV), are emerging as a flexible and cost-effective solution to provide on-demand wireless connectivity in such scenarios. UAVs have the possibility to operate virtually everywhere, and the growing payload capacity makes them suitable platforms to carry wireless communications hardware, playing the role of mobile base stations, access points or relay nodes. A flying network may typically be composed of a fleet of UAVs, organized in a multi-tier topology with so-called Flying Edge Nodes (FENs) and Flying Gateways (FGWs)~\cite{COELHO2023103000}. Figure~\ref{fig:flyingNetwork} shows a flying network example where FENs can play the role of Flying Access Points that provide the access network to the users on the ground, or the role of Flying Sensor Nodes that can perform video surveillance missions. The FENs forward the traffic to the FGWs, that act as relay nodes and are responsible for forwarding the traffic to/from the backhaul (BKH) network and ultimately to/from the Internet. 

Most of the state of the art works propose rate adaptation solutions that do not consider the specific characteristics of flying and vehicular networks~\cite{Yin_2020}. In flying networks, the nodes need to be properly positioned and their wireless link configuration dynamically adjusted in order to ensure the Quality of Service (QoS) expected by the end users. In addition, these scenarios are typically highly unpredictable due to the varying locations as well as the concentration/dispersion of end-users served by the Flying Access Points and their movements regarding direction and velocity - e.g., vehicles or pedestrians. Therefore, a static wireless link configuration and UAV positioning are not adequate. 

The ResponDrone project~\cite{respondrone} defined different flying network use cases (e.g., Follow-me missions) where the future trajectory of flying nodes is known, as requested by the mission commander. However, the usage of this information to predict future wireless channel conditions has been overlooked. State of the art work has been mainly focused on the optimal positioning of the flying nodes, having most of the wireless link parameters statically configured with default values. The Rate Adaptation challenge is well-known in fixed or low mobility IEEE 802.11 networks, and Minstrel High Throughput (HT)~\cite{minstrelht} is the default Wi-Fi rate adaptation algorithm used in the Linux kernel since the IEEE 802.11n version. Yet, it performs inefficient random sampling of the environment and reacts with significant delays in situations where the link quality improves~\cite{smartla}. The authors in~\cite{He_2019,Xiao_2021} consider information of UAV sensors to estimate wireless channel conditions for unknown environments and perform rate adaptation, but they do not use the information about the UAV trajectories as input. To the best of our knowledge, solutions that use the node trajectory information to predict the wireless channel conditions and perform rate adaptation are yet to be developed.

The main contribution of this paper is the Trajectory-Aware Rate Adaptation (TARA) approach. TARA takes advantage of knowing the trajectory of all nodes in the flying network to estimate future changes in the wireless link quality and perform rate adaptation accordingly. The network performance improvement achieved with TARA was evaluated using ns-3~\cite{ns3}. The simulation results show significant throughput gains when compared with conventional rate adaptation algorithms.

The rest of the paper is organized as follows. Section II explains the TARA approach and its implementation. Section III evaluates TARA by means of simulation. Finally, Section IV provides some concluding remarks and points out the future work.

\section{Trajectory-Aware Rate Adaptation Approach}

TARA is a standard-compliant solution that builds on top of Minstrel-HT and aims at overcoming the Minstrel-HT limitations by taking advantage of the knowledge about the future movement of the flying nodes. It predicts the future Modulation and Coding Scheme (MCS) index to use ($MCS_{TARA}$) and changes the Minstrel-HT retry chain table to consider the predicted $MCS_{TARA}$. 

The problem involves the dynamic data rate optimization to maximize the throughput of a flying network composed of FENs, one or more FGWs and the respective BKH link(s). The movement of a particular FEN is determined by its mission, which can be either predefined or adjusted based on specific objectives during the mission. In order to react to the movements of FENs, the FGWs must move to new positions that must be calculated based on one or more criteria so that the network performance is optimized. 

\begin{figure}
  \centering
  \includegraphics[trim={0 0.325cm 0 0cm},clip,width=.9\linewidth]{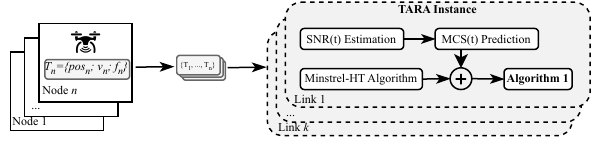}
  \caption{Illustration of the TARA approach.}
  \label{fig:TARA-Architecture}
\end{figure}

The TARA approach is depicted in Figure~\ref{fig:TARA-Architecture}. Table~\ref{tab:notations} defines the notation used hereafter. TARA has an instance associated to each wireless link and uses the trajectory information of every node in the flying network as input for the $SNR(t)$ estimation function. This function is then used to predict the $MCS(t)$ value that is applied in an improved version of the Minstrel-HT algorithm. Despite its design being based on IEEE 802.11 and related rate adaptation algorithms, with the proper adjustments, TARA can be applied to other wireless communication technologies. 

\begin{table}
  \centering
  \caption{Defined notations}
  \label{tab:notations}
  \begin{tabular}{cl}
    \hline
    Notation & Description \\
    \hline
    $T_n$             & $n$th node trajectory                      \\
    $pos_n$           & $n$th node current position: $(x,y,z) m$   \\
    $\vec{v_n}$       & $n$th node velocity: $(v_x,v_y,v_z) m/s$   \\
    $f_n$             & $n$th node flight duration in seconds      \\
    $SNR_k$           & $k$th link Signal to Noise Ratio           \\ 
    $SNR_{threshold}$ & Min. SNR value for target (BER, MCS)       \\
    $MCS_{TARA}$      & TARA MCS prediction for time interval $t$  \\
    $MCS_{MaxTP}$     & Minstrel-HT best throughput MCS            \\
    $MCS_{MaxTP2}$    & Minstrel-HT 2nd best throughput MCS        \\
    $MCS_{MaxProb}$   & Minstrel-HT best probability MCS           \\
    $TX_{MCS}$        & Minstrel-HT current transmission MCS       \\
    \hline
\end{tabular}
\end{table}

In what follows, the movement of a flying node is modeled and an estimation method of wireless link quality is proposed, based on trajectory information. Next, the prediction of $MCS_{TARA}$ is explained; it determines the PHY rate for the next frame transmission(s) in the respective link. Finally, the improvement of the Minstrel-HT rate adaptation algorithm is detailed.

\subsection{SNR estimation based on node trajectories}

The trajectory of a node $n$ is defined as the data tuple $T_n=\{pos_n; \vec{v_n}; f_n\}$, as shown in Figure~\ref{fig:TARA-Architecture}, where $pos_n$ is the current position of the node, $\vec{v_n}$ is the node velocity and $f_n$ is the node flying time. With the trajectory information we can calculate the position function for any node in the network as $pos_n+\vec{v_n}t$ as long as $t \leq f_n$. This is because, by then, the node has stopped flying and its position does not change.

The communication channel between two flying nodes, characterized by a strong line of sight component, is modeled using the Friis path loss model~\cite{channel-uavs}. Considering a link $k$, which connects nodes $a$ and $b$, $SNR_k(t)$ in dB is given by: \begin{equation}
  SNR_k(t)=P_{t}^{dB} + G_{t}^{dBi} + G_{r}^{dBi} - P_{n}^{dB} - FSPL^{dB}(t)
  \label{eq:SNR}
  \end{equation}
  \begin{equation}
    FSPL^{dB}(t)= 20\log_{10}\left( \frac{4\pi}{\lambda}d(pos_a(t),pos_b(t))\right)
    \label{eq:FSPL}
  \end{equation} where $d(pos_a(t),pos_b(t))$ is the Euclidean distance between nodes $a$ and $b$. Note that the computed $SNR_{k}$ values assume a discretized $t$ that is synchronized with the update frequency of the rate adaptation algorithm; this is detailed in the following sections.

\subsection{Calculation of MCS predicted value}\label{sec:mcs-pred}
The optimal MCS to use in a wireless communications link is often a complex task due to unpredictable factors -- other than SNR -- that impact the channel conditions, such as interference, shadowing and multipath fading, to name a few. In \cite{nistval}, emulation trials in a physical-layer testbed showed significant differences from ns-3 simulation results. Motivated by this, in \cite{pei2011validation} the \emph{NistErrorRateModel} is proposed to model orthogonal frequency division multiplexing transmissions with generally $\pm1~dB$ margin when compared with emulation data, in interference-free scenarios.

Using \emph{NistErrorRateModel}, we define a Bit Error Ratio (BER), $\rho_{BER}$. Then, a look-up table that maps $(MCS_i,\rho_{BER})\mapsto SNR_{threshold}$ is built. $SNR_{threshold}$ is the minimum SNR value that is required to achieve the required BER for the specified MCS, and $i$ is the MCS index. 

The output of Eq.~\ref{eq:SNR}, $SNR_{k}$, is used to choose the best achievable $MCS_i$, given the previously calculated $SNR_{threshold}$, as long as the condition $SNR_{k}\geq SNR_{threshold}$ is verified. Herein, we refer to the best achievable $MCS_i$ as $MCS_{TARA}$, which is used to modify the original Minstrel-HT algorithm, as shown in Figure~\ref{fig:TARA-Architecture}.

\subsection{Minstrel-HT Algorithm Improvement} \label{sec:ra}

Minstrel-HT is the evolution of the Minstrel algorithm~\cite{ra-surv,minstrel-eval}, which incorporates new IEEE 802.11n features such as the configuration options of Number of Spatial Streams (NSS), Guard Interval (GI) and Channel Bandwidth (CB). In~\cite{minstrel-eval, smartla} the authors observe that both Minstrel and Minstrel-HT underperform in scenarios with dynamic channel conditions. 

To overcome the problem, we propose a novel version of Minstrel-HT to improve its overall performance in dynamic scenarios such as Flying Networks. Algorithm~\ref{alg:tara} highlights the main modifications that were implemented so that the original Minstrel-HT algorithm uses the $MCS_{TARA}$, without compromising its original operation, in situations where the $MCS_{TARA}$ calculation is inaccurate, and the retry chain table is necessary. This inaccuracy, expressed by the need of frame retransmissions, is due to the selection of an optimistic MCS value. The existing functions that were modified are detailed in the following.

\begin{algorithm}
\caption{Improvement of Minstrel-HT with trajectory-aware MCS prediction, $MCS_{TARA}$.}
\label{alg:tara}
\begin{algorithmic}[1]
\Procedure{UpdateStats}{$link$}\Comment{from Minstrel-HT}
  \State ... \Comment current algorithm
  \If{$MCS_{TARA} > MCS_{MaxTP}$}
    \State $MCS_{MaxTP2} \gets MCS_{MaxTP}$
    \State $MCS_{MaxTP} \gets MCS_{TARA}$
  \EndIf
\EndProcedure
\Procedure{UpdateRate}{$link$}\Comment{from Minstrel-HT}
  \If{$retries < 2$}
    \State $TX_{MCS} \gets MCS_{TARA}$
  \ElsIf{$retries < RetryCount_{MaxTP}$}
    \State $TX_{MCS} \gets MCS_{MaxTP}$
  \ElsIf{$retries < RetryCount_{MaxTP2}$}
    \State $TX_{MCS} \gets MCS_{MaxTP2}$
  \ElsIf{$retries < RetryCount_{MaxProb}$}
    \State $TX_{MCS} \gets MCS_{MaxProb}$
  \EndIf
\EndProcedure
\end{algorithmic} 
\end{algorithm}

\subsubsection{UpdateStats} It is called every $\tau$ milliseconds to update the link-specific statistics. These include the number of successes and attempts for each MCS in each $(NSS, GI, CB)$ group, which impact both the expected throughput and exponentially weighted moving average probability of success of each MCS. Finally, the decision of Minstrel-HT's MCS $(MCS_{MaxTP}, MCS_{MaxTP2}, MCS_{MaxProb})$ is updated, for the following $\tau$ period.

$MCS_{TARA}$ is compared with $MCS_{MaxTP}$. If $MCS_{TARA}$ is higher than the current best throughput MCS, the Minstrel-HT's MCS is updated to $(MCS_{TARA}, MCS_{MaxTP}, MCS_{MaxProb})$, in that specific order. Otherwise, the TARA suggestion is used to modify the original retry chain table; this is explained next.

\subsubsection{UpdateRate} It implements the retry chain table. Considering that a transmitted frame is lost for any reason, Minstrel-HT will attempt to retransmit that lost frame, with the previously used MCS, or any other in the retry chain table, until the maximum number of $retries$ is reached. Minstrel-HT calculates the amount of retries ($RetryCount$) for each of its MCS values. For $MCS_{TARA}$, a fixed amount of 2 retries is defined, before resuming the original Minstrel-HT retry chain table. In this way, we impose the trajectory-aware MCS suggestion without disrupting the original operation of the algorithm, since after 2 retries -- 3 frame transmissions attempts -- the algorithm falls back to its original behavior. 

\section{Performance Evaluation}

The flying network performance achieved with TARA is presented in this section, including the simulation scenario, the simulation setup and the analysis of results.

\subsection{Simulation Scenario}\label{sec:sims_cen}

Herein, we aim to evaluate the performance of the new proposed Rate Adaptation algorithm; for this purpose, a simple simulation scenario with a single FEN and FGW is sufficient. This choice allows for a preliminary evaluation of the TARA feasibility and provides valuable insights for future work involving more complex scenarios with multiple FENs.

In this scenario, the BKH and the access links must carry the same amount of traffic, except for possible losses that should be kept as low as possible. For that reason, the position of the FGW was assumed to remain in the midpoint between the BKH and the FEN. In this way, it equalizes the SNR value on both links, which is only affected by the distance, and thus the same MCS index. However, such an assumption is not generalizable to scenarios with multiple FENs with heterogeneous traffic demand.

Extending the problem to scenarios with multiple FENs will be part of future work, which will include the development of a novel positioning algorithm for the FGWs, aimed at optimizing the system performance, while relying on the TARA approach.

\subsection{Simulation Setup}

In order to evaluate the flying network performance achieved with TARA, ns-3 (version ns-3.38) was used. A summary of the most relevant configuration parameters is presented in Table~\ref{tab:sim-params}. Different Wireless Local Area Network (WLAN) channels were used for each link, to ensure that there are no frame losses due to interference. Traffic was generated using User Datagram Protocol (UDP), with a constant packet size of 1400~bytes and a data rate above link capacity, to saturate the communication link. The TARA performance was compared with the original Minstrel-HT and Ideal~\cite{ns3}. 

The Ideal algorithm implementation maintains for every link the SNR value of every packet received and sends back this SNR value to the sender by means of an out-of-band mechanism. Each sender keeps track of the last SNR value sent back by a receiver and uses it to pick a transmission mode based on a set of SNR thresholds derived from a target BER, and transmission mode-specific SNR/BER curves, as it was explained in Section~\ref{sec:mcs-pred}.

\begin{table}
  \centering
  \caption{ns-3 Simulation Parameters}
  \label{tab:sim-params}
  \begin{tabular}{ll}
  \hline
  Configuration Parameter & Value \\
  \hline
  Wi-Fi Standard          & IEEE 802.11n                      \\
  Propagation Delay Model & Constant Speed                    \\
  Propagation Loss Model  & Friis                             \\
  Error Rate Model        & NistErrorRateModel                \\
  Channel Bandwidth       & 20 MHz                            \\
  Transmission Power      & 20 dBm                            \\
  RX/TX antenna gains     & 0 dBi                             \\
  Wi-Fi MAC               & Ad-hoc                            \\
  $\rho_{BER}$            & 1e-6                              \\
  $\tau$                  & 50 milliseconds                   \\
  $\Delta$                & 30 seconds                        \\
  \hline
\end{tabular}
\end{table}

\begin{figure}
  \centering
  \includegraphics[trim={0 0.375cm 0 0.25cm},clip,width=.8\linewidth]{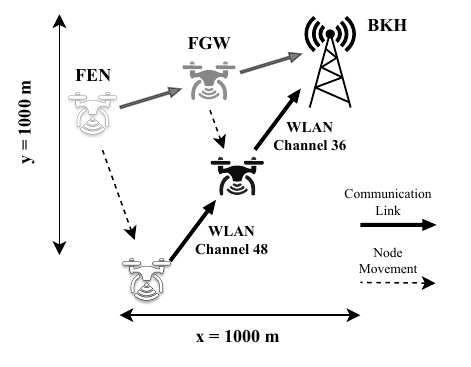}
  \caption{Network topology. The communication links are represented by full lines, with two non-interfering WLAN channels.}
  \label{fig:net-topology}
\end{figure}

The dynamic network topology is represented in Figure~\ref{fig:net-topology}; uplink traffic generation by the FEN is relayed by the FGW, with BKH as destination. Random FEN movements were defined for each simulation run using different seeds, which provide a broad and rich sequence of independent events~\cite{random_trajectories} that put stress on the Rate Adaptation algorithm, in order to evaluate its performance. A complete trajectory (for the duration of a simulation run) is a sequence of elementary movements along a linear path. Every simulation run starts with a new random position of the FEN and its trajectory is randomly updated every $\Delta$ seconds. Each movement starts at the final position of the previous one, and a new path is randomly generated (direction and length). When moving, the FEN velocity remains constant at 8~m/s. The duration of each movement is $t_m \leq \Delta$. For the FGW a different approach was used. As said, the final position of each FGW movement (along a linear path, as well) must be at the midpoint of the straight line between the BKH and the final position of the corresponding FEN movement. Moreover, the FGW moves with a constant velocity, such that both the FEN and the FGW arrive at the same time to their final positions. By imposing these conditions, all the intermediate FGWs positions have the same property. The BKH node is fixed at the edge of the scenario, which is a square with 1000 by 1000~m.

We defined simulation runs with a duration of 300 seconds, which were evaluated using 100 different random seeds. These random seeds impact the initial position of the FEN, the sequence of random elementary movements (paths) as well as the average values of throughput. The results are expressed using mean values of the throughput measured at the MAC layer every simulation second, and their relative percentage gains. The measured throughput is link-specific. We present results for both the relay link (between the FGW and BKH node) and the access link (between the FEN and the FGW). However, the end-to-end network performance is best characterized by the relay link since the effective throughput of the system is determined by the packets delivered to the BKH after being forwarded by the FGW.

\subsection{Simulation Results}

Simulation results are presented in this section, first those referring to a particular random scenario (seed) and then results of the extensive simulations. Finally, a discussion of the results is provided.

\begin{figure*}
  \centering
  \subfloat[Distance between nodes throughout the simulation period.]{
      \includegraphics[width=.3\textwidth]{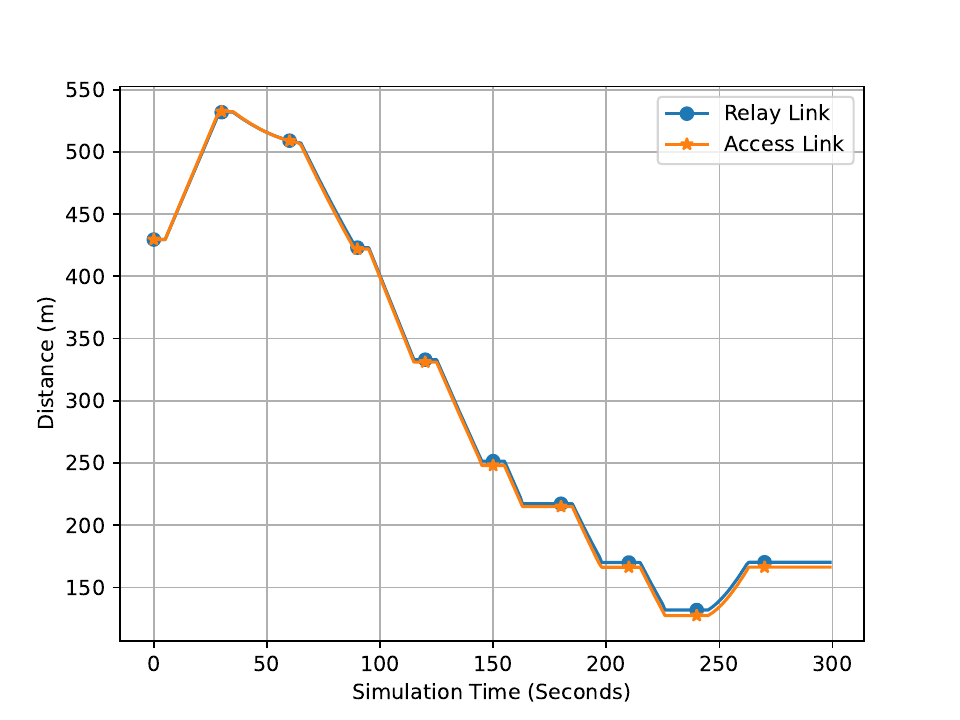}
      \label{fig:73dis}
  }
  \hfil
  \subfloat[Relay link Throughput, throughout the simulation period.]{
      \includegraphics[width=.3\textwidth]{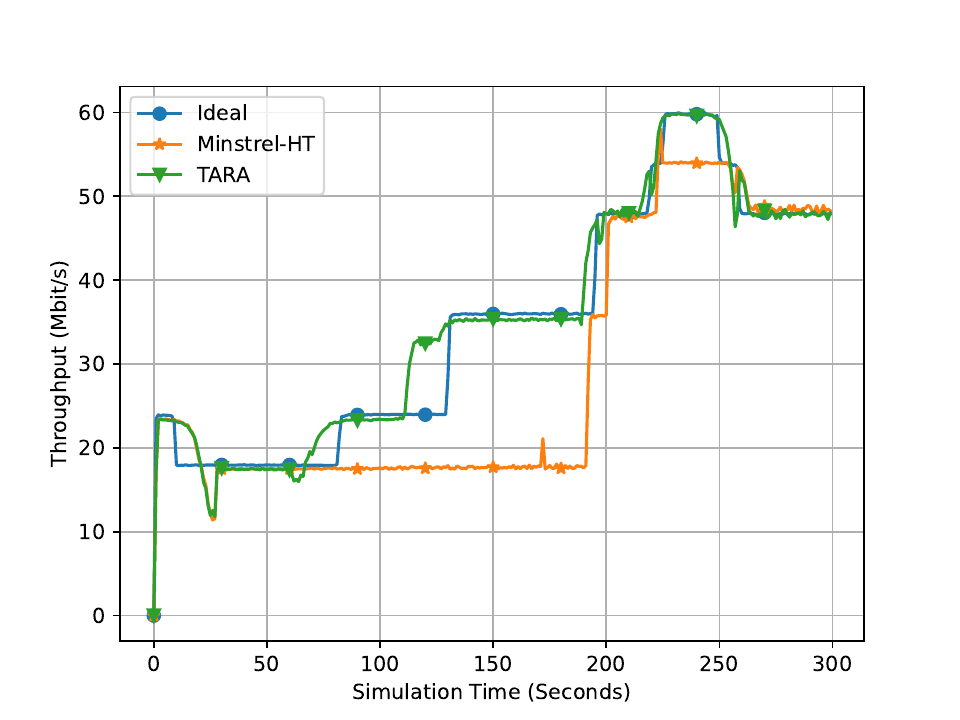}
      \label{fig:73thp}
  }
  \hfil
  \subfloat[Relay link Throughput CCDF.]{
      \includegraphics[width=.3\textwidth]{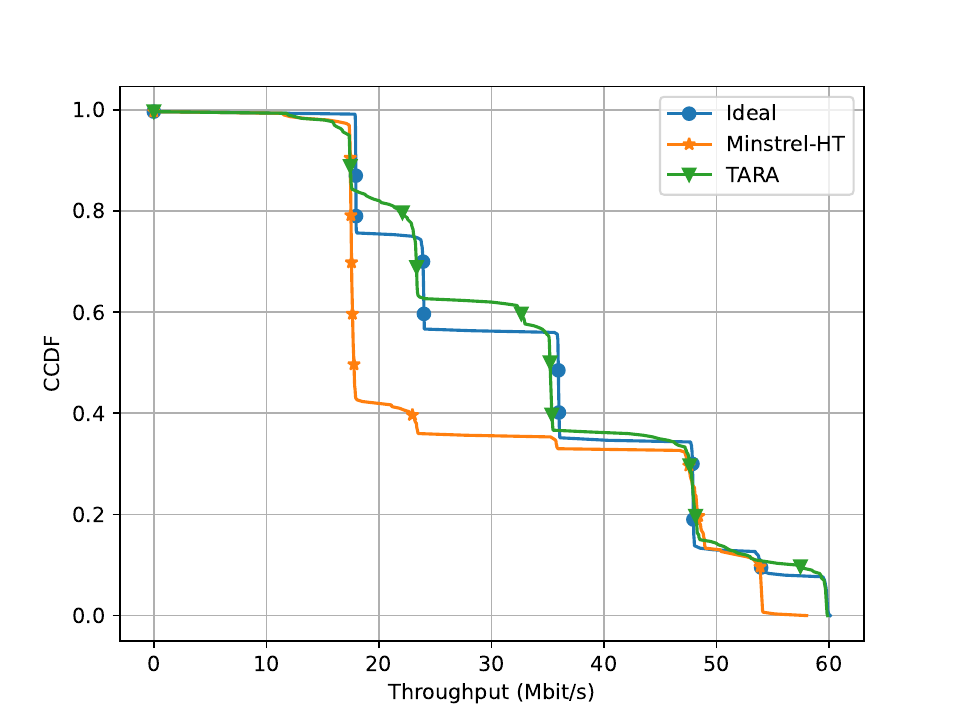}
      \label{fig:73ccdf-thp}
  }
  \caption{Selected Seed Results.}
\end{figure*}

\subsubsection{Example Scenario}

A random seed was selected, which generated a simulation scenario covering a wide range of distances, as shown in Figure~\ref{fig:73dis}. The throughput achieved in the relay link measured throughout the simulation period is shown in Figure~\ref{fig:73thp}. As expected, the throughput increases as the distance between nodes in each link decreases, since the SNR value is higher, which makes the use of higher MCS indexes possible. We can also observe that TARA, despite being based on Minstrel-HT, changes MCS indexes with a faster reaction time, which can be directly compared with the reaction time of the Ideal algorithm. In Figure~\ref{fig:73ccdf-thp} we represent the Complementary Cumulative Distribution Function (CCDF) of the throughput for the same seed. The CCDF $F(x)$ represents the percentage of time for which the mean throughput was higher than $x$. From the results we observe the significant throughput gains of TARA.

The reason for slight improvement of TARA over the ``not so" Ideal algorithm, is because the Ideal algorithm is optimized towards minimizing the BER, while the results are comparing the solutions in terms of throughput. With these results we can observe that TARA's slight gains are due to few decision instants, where specific conditions are met. These conditions are observed when the decision of adapting MCS could result in a higher BER probability, but the throughput would still be better.   

The corresponding results for the access link are not presented since they were similar due to the fact that both links have the same distance, which optimizes performance. However, these results are heavily biased by the initial position of the nodes and their trajectory. To address the impact of different random trajectories, we present in the following section the results for 100 different random seeds.

\subsubsection{Extensive simulations}

\begin{figure}
  \centering
  \includegraphics[trim={0 1cm 0 6.6cm},clip,width=.85\linewidth]{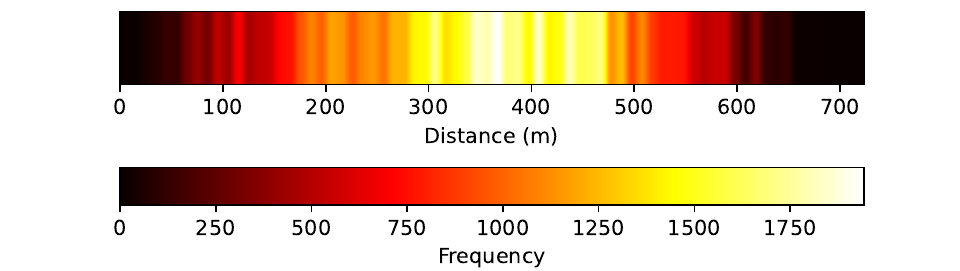}
  \caption{Link distance distribution, with $3*10^4$ total samples.}
  \label{fig:dis_heat}
\end{figure}

Figure~\ref{fig:dis_heat} represents the distribution of distances that were observed during the extensive simulations that were carried out. The higher frequency of distances between 300 and 500 meters can help justify the distribution of the observed distribution of throughput values as well.

\begin{figure*}
  \centering
  \subfloat[Throughput CCDF for the Relay Link, between the BKH and the FGW.]{
      \includegraphics[width=.3\textwidth]{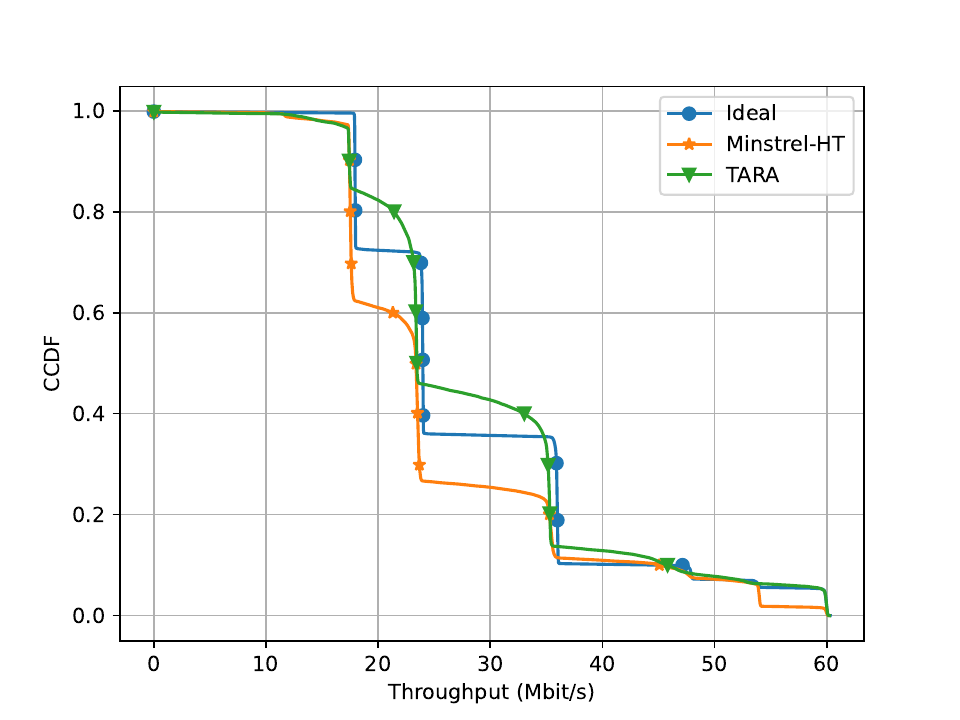}
      \label{fig:ccdf-relay}
  }
  \hfil
  \subfloat[Mean Throughput with 99\% confidence interval.]{
      \includegraphics[width=.3\textwidth]{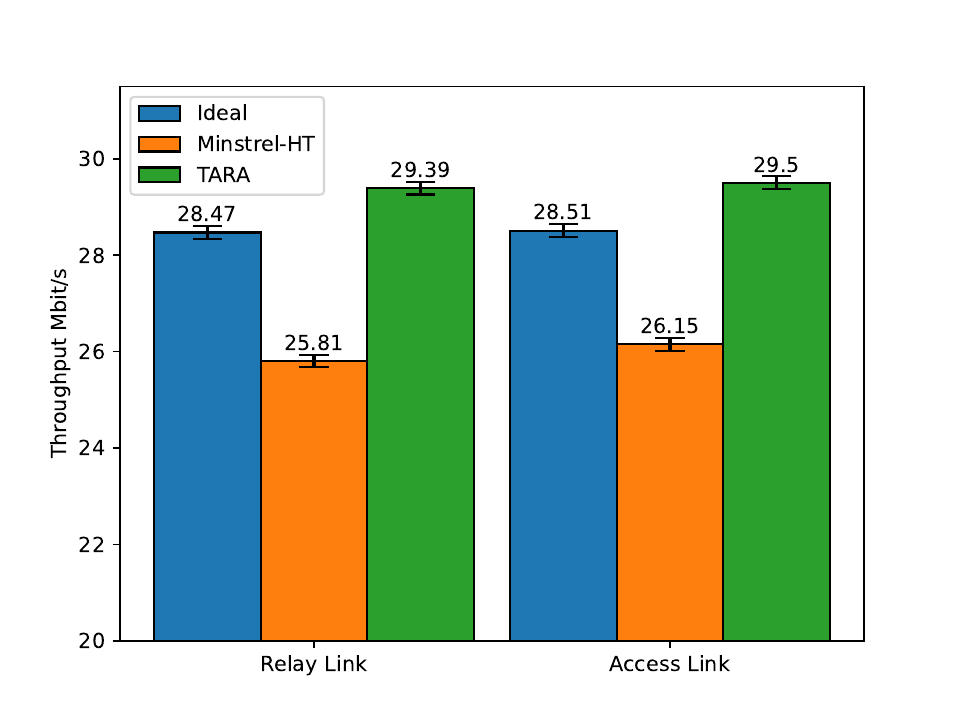}
      \label{fig:ci}
  }
  \hfil
  \subfloat[Throughput CCDF Gains for both links, comparing TARA with Minstrel-HT and Ideal rate adaptation algorithms.]{
      \includegraphics[width=.3\textwidth]{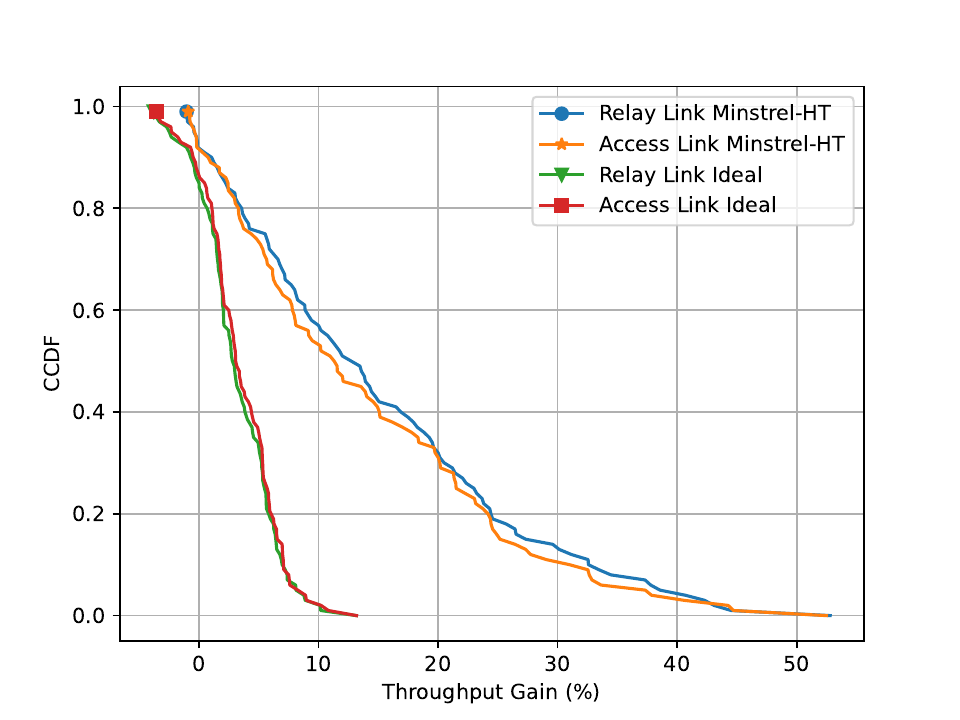}
      \label{fig:ccdf-gains}
  }
  \caption{Extensive simulation results.}
\end{figure*}

Figure~\ref{fig:ccdf-relay} represents the CCDF of the throughput for the relay link. In this analysis three different percentiles, $70^{th}$, $50^{th}$ and $30^{th}$ are considered. For the $30^{th}$ percentile, there is a throughput improvement of 31.5\% when compared to Minstrel-HT and a throughput deterioration of 2.7\% when compared to Ideal. If we consider the $50^{th}$ percentile (median), the throughput results are all within 2\% difference of each other. However, for the $70^{th}$ percentile, there is a throughput improvement of 48.4\% when compared with Minstrel-HT and a throughput deterioration of 2.1\% when compared with Ideal.

Figure~\ref{fig:ci} represents the 99\% confidence interval of the mean throughput for both links. Regarding the relay link, the mean throughput increases by 13.9\% and by 3.2\% when compared with Minstrel-HT and Ideal respectively. Regarding the access link, the performance is similar to the relay link.

Figure~\ref{fig:ccdf-gains} represents the CCDF of the percentage gains of TARA throughput when compared with Minstrel-HT and Ideal. The gains were calculated considering the mean throughput of each random seed. The CCDFs are similar for each link, thus its analysis will be made considering solely the relay link results, since it better represents the end-to-end network performance. Positive throughput gains of TARA relatively to Minstrel-HT and Ideal occurred in 92\% and 86\% of the seeds, respectively. 
The highest throughput gain of TARA, on a particular seed, was 52.8\% and 13.2\%, when comparing with Minstrel-HT and Ideal, respectively.

\subsection{Discussion}

In this section we address the observations regarding the average throughput values in the access link and relay link, the comparison between TARA and Ideal throughput, and the significance of the 14\% average throughput gain over Minstrel-HT algorithm:

\begin{itemize}
  \item The average throughput in the access link was slightly higher than in the relay link. This is an expected result since frame loss can occur in both access and relay link.
  \item When comparing TARA and Ideal throughput, it might be surprising that the average throughput of TARA is 3.2\% higher than that of Ideal. However, upon analyzing the CCDF curves, the TARA curve either coincides or lies above the Ideal curve. This finding explains the higher average throughput of TARA. Despite the lower percentiles indicating potential discrepancies, the overall behavior of the CCDF curves suggests that TARA achieves comparable or superior performance to the Ideal algorithm, leading to the higher average throughput.
  \item Note that a 14\% gain over Minstrel may seem modest. However, it is important to consider the stability of TARA's performance and the potential for greater gains when the algorithm needs to adapt to frequent improvements in the connection quality. The selected seed represents a case where successive increases in the modulation and coding scheme (MCS) were necessary, resulting in a throughput percentage gain of 21.2\% when compared with Minstrel-HT. These findings suggest that TARA's stability and adaptability provide benefits beyond a simple percentage gain, and the actual gains can be more significant in scenarios with frequent improvements in the link quality.
\end{itemize}

In summary, the results demonstrate that TARA matches the performance of the Ideal algorithm. This is relevant because the Ideal algorithm can be considered as a benchmark algorithm; it is not implementable due to its lack of standard compliance. Despite the 14\% gain over Minstrel-HT, the stability and adaptability of TARA make it a promising algorithm, particularly in situations where frequent enhancements to link quality occur. Further investigation into the extreme case with successive MCS increases could provide additional insights into TARA's performance.

\section{Conclusions and Future Work} \label{sec:conclusions}

This paper proposes TARA, a trajectory-aware rate adaptation approach for flying networks. TARA takes advantage of the knowledge of future movements of UAVs to predict how the quality of wireless links will change, and perform rate adaptation accordingly. The proposed solution was evaluated using ns-3 simulations. Simulation results showed consistent gains when compared with state of the art algorithms,  such as Minstrel-HT, with a throughput increase of up to 53\% for the simulated scenarios. The results for each seed and the TARA source code\footnote[2]{https://gitlab.inesctec.pt/pub/ctm-win/tara} are publicly available~\cite{tara-dataset}. 

As future work, we plan to evaluate TARA experimentally, compare it with other state-of-the-art algorithms and address the complexity of scenarios with multiple FENs and more than one FGW. These challenges include achieving an equal throughput on both sides of the FGW, considering the trade-off between distance and throughput in multi-link scenarios, handling cases where optimal positioning cannot guarantee target throughput on all links, and accounting for variable and time-dependent traffic characteristics. Finally, we aim at improving TARA to take into consideration stochastic path loss models, scenarios with communication interference, and drone movements that consider the inertia that exists in real world systems.  

\section*{Acknowledgments}

This work is financed by National Funds through the Portuguese funding agency, FCT - Fundação para a Ciência e Tecnologia, under the PhD grant 2022.10093.BD.

\bibliographystyle{IEEEtran}
\bibliography{refs}

\end{document}